\documentclass[11pt]{article}
\usepackage[titletoc]{appendix}
\usepackage[labelsep=period]{caption}
\usepackage{bm}
\usepackage{fullpage}
\usepackage{epsfig}
\usepackage{graphics}
\usepackage{latexsym}
\usepackage{amsmath}
\usepackage{amsfonts}
\usepackage{amssymb}
\usepackage{mathrsfs}
\usepackage{pifont}
\usepackage{yhmath}
\usepackage{fullpage,graphicx}
\usepackage{enumitem}
\usepackage{subcaption}
\usepackage{float}
\usepackage{mathtools}

\usepackage{bbm}
\usepackage{dsfont}
\usepackage{eufrak}
\usepackage{wasysym}
\usepackage{underscore}
\usepackage{epstopdf}
\usepackage{fontenc}
\usepackage{amsthm}

\newcommand{\de}{\backslash}
\DeclareMathAlphabet{\mathpzc}{OT1}{pzc}{m}{it}
\title{\bf How Difficult Is It to Recognize CIS Graphs?}
\author{\vspace{2mm}  Rongchuan Tao$^{a}$ \quad Mengxi Yang$^{b}$\thanks{Corresponding author. E-mail: yangmx221b@outlook.com.}
\quad Wenan Zang$^{a}$\thanks{Supported in part by the Research Grants Council of Hong Kong.}\\
$\stackrel{a}{}$ Department of Mathematics\\University of Hong Kong \\ Hong Kong, China\smallskip\\
$\stackrel{b}{}$ School of Mathematical Sciences\\ University of Science and Technology of China\\Hefei 230026, China}

\begin{document}
\date{}
\maketitle

\begin{abstract}
A graph $G$ is called {\em CIS} if each maximal clique intersects each maximal stable set of $G$, with maximality taken with respect to set inclusion. CIS graphs 
resemble perfect graphs in several respects and have interesting applications in game theory. The complexity of recognizing CIS graphs was posed as an open 
problem by Chv\'atal in the 1990s and has since led to conflicting conjectures. We settle the problem by showing that recognizing CIS graphs is 
\(\mathsf{coNP}\text{-complete}\).

\vskip 4mm


\noindent {\bf OR/MS subject classification.} Primary: Programming/graphs.


\end{abstract}

\newpage

\section{Introduction}

A classical theorem in combinatorics, due to Grillet \cite{G}, asserts that in every partially ordered set containing no 
quadruple $(a,b,c,d)$ such that $a<b$, $c<d$, $b$ covers $c$, and the remaining three pairs of elements are incomparable,
each maximal chain meets each maximal antichain, where (and throughout) the adjective {\em maximal} is meant with 
respect to set inclusion rather than size. Over the past three decades, various attempts have been made to generalize this 
theorem. A graph $G$ is called {\em CIS} if each maximal clique intersects each maximal stable set of $G$. The graphs called CIS 
here are exactly {\em Grillet} graphs introduced by Chv\'atal in \cite{Ch,Za}. Recall that a {\em clique} is a set of pairwise adjacent 
vertices, while a {\em stable set} is a set of pairwise nonadjacent vertices. As observed by Andrade, Boros, 
and Gurvich \cite{ABG,ABG2}, CIS graphs also have interesting applications in game theory. In \cite{B}, Berge made a conjecture and 
suggested a research problem concerning CIS graphs, both of which were resolved by Zang \cite{Za}. Let $P_4$ denote an 
induced path with four vertices $v_1, v_2, v_3, v_4$, and let $A$ be obtained from $P_4$ by introducing a fifth vertex $v_5$ and 
making it adjacent to $v_2$ and $v_3$. Let $F$ be obtained from a cycle $u_1u_2u_3u_4u_5u_6u_1$ by adding a triangle $u_1u_3u_5u_1$. 
Chv\'atal proposed the following conjecture \cite{Ch,DLZ} as a variation on Berge's problem: Let $G$ be a graph with no induced subgraph 
isomorphic to $F$ or its complement. Then $G$ is CIS if and only if each $P_4$ extends to an $A$ in $G$. This conjecture was 
confirmed by Deng, Li, and Zang \cite{DLZ,DLZ2}. Independently, Andrade, Boros, and Gurvich \cite{ABG,ABG2} formulated the same statement 
and proved it using a different approach. For further work on CIS graphs, we refer the reader to \cite{AGM,BGM,BGZ,DHMV,WZZ}; for comprehensive 
accounts of CIS graphs, see \cite{ABG,ABG2}.

Chv\'atal \cite{Ch,Za} also suggested the following problem in the 1990s. 

\vskip 2mm
\noindent {\bf Problem}. {\em How difficult is it to recognize CIS graphs?}
\vskip 2mm 

The complexity of this recognition problem has been the subject of sharply differing views. 

On the one hand, if a graph $G$ is not CIS, then this fact cannot be certified by exhibiting a forbidden induced subgraph, because 
every graph is an induced subgraph of some CIS graph. To see this, let $C_1,C_2,\ldots,C_k$ be all maximal cliques of $G$. Add to $G$ 
pairwise nonadjacent vertices $v_1,v_2,\ldots,v_k$, and join $v_i$ to every vertex of $C_i$ for each $1\le i\le k$. Moreover, Zang \cite{Za} 
showed that, given a graph $G$ together with a specified maximal stable set $S$, it is \(\mathsf{coNP}\text{-complete}\) to decide whether $S$ intersects 
every maximal clique of $G$. Since the CIS property is not closed under taking induced subgraphs, and since a graph may have exponentially 
many maximal cliques or maximal stable sets, the recognition problem has been conjectured to be \(\mathsf{coNP}\text{-complete}\); see, for instance,
Zverovich and Zverovich \cite{ZZ} and Zang \cite{Za}.

On the other hand, CIS graphs resemble perfect graphs \cite{CRST} in several respects. For example, the class of CIS graphs is closed under
taking complements, neighborhoods, and non-neighborhoods. It can also be shown that if a graph $G$ is obtained from the disjoint
union of two graphs $H$ and $K$ by deleting a vertex $v$ of $H$ and then adding all possible edges between $K$ and the neighbors of
$v$ in $H$, then $G$ is CIS if and only if both $H$ and $K$ are CIS. In view of these favorable structural properties, Andrade,
Boros, and Gurvich \cite{ABG} conjectured that recognizing CIS graphs is solvable in polynomial time, albeit with considerable
difficulty.

Thus, substantial evidence has been offered on both sides, suggesting that the problem lies near the boundary between polynomial-time
solvability and computational intractability. This makes Chv\'atal's problem particularly intriguing.

The purpose of this note is to establish the following complexity result.

\vskip 2mm
\noindent {\bf Theorem}. {\em Recognizing CIS graphs is \(\mathsf{coNP}\text{-complete}\).}

\section{Proof}

Clearly, the recognition problem belongs to \(\mathsf{coNP}\). To establish the assertion, it suffices to reduce the $3$-{\small SATISFIABILITY} 
problem (\(\mathsf{3SAT}\)) \cite{GJ} to the complement of this problem. Let ${\cal C}=\{c_1,c_2,\ldots,c_m\}$ be the set of clauses 
in an instance of \(\mathsf{3SAT}\) given in conjunctive normal form (\(\mathsf{CNF}\)), where each clause contains precisely three distinct 
literals. We shall construct a graph $G=(V,E)$, such that $G$ is not CIS if and only if ${\cal C}$ is satisfiable. 

Let $U=\{u_1,u_2,\ldots,u_n\}$ be the set of variables appearing in ${\cal C}$. We may assume, without loss of generality, the following.

(1) Each literal $\sigma \in \cup_{i=1}^n \{u_i, {\bar u}_i\}$ occurs in at least one clause. Otherwise, setting $\bar{\sigma}$ to {\em true} would
satisfy every clause containing $\bar{\sigma}$, thereby simplifying the given instance of \(\mathsf{3SAT}\). 

(2) No literal $\sigma \in \cup_{i=1}^n \{u_i, {\bar u}_i\}$ occurs in every clause. Otherwise, setting $\sigma$ to {\em true} would immediately 
satisfy the instance, making it trivial.  

(3) No clause contains both $u_i$ and ${\bar u}_i$ for any $1\le i \le n$. Otherwise, this clause would be satisfied by every truth assignment and 
hence could be removed from the instance. 

(4) $n \ge 4$. Otherwise, we can add two new variables $u_{n+1}$, $u_{n+2}$ and two new clauses $c_{m+1}=(u_1 \vee u_{n+1} \vee u_{n+2})$, 
$c_{m+2}=(u_2 \vee {\bar u}_{n+1} \vee {\bar u}_{n+2})$ to the instance without changing its satisfiability.    

For each $\sigma \in \{u_i, {\bar u}_i\}$, let $\sigma^*:=u_i$ denote the variable corresponding to $\sigma$. For each $c_j \in {\cal C}$, define 
$Z_j:=\{u_k: \mbox{neither $u_k$ nor ${\bar u}_k$ occurs in} \,\, c_j \}$. By (4), we have $Z_j \ne \emptyset$.

The graph $G$ is constructed as follows: 

\begin{enumerate}[leftmargin=36pt]
\vspace{-2mm}
\item[(5)] {\hspace{-1.5pt}}{For each variable $u_i\in U$, introduce a truth-setting component $F_i$ consisting only of the edge $u_i{\bar u}_i$.
Note that these components are pairwise disjoint. With a slight abuse of notation, each literal of ${\cal C}$ also denotes
the corresponding vertex of $G$.}  
\vspace{-2mm}
\item[(6)] {\hspace{-1.5pt}}{For each clause $c_j\in {\cal C}$, introduce a satisfaction-testing component $H_j$ consisting of an isolated 
vertex $c_j$ and pairwise disjoint edges $u_k^j{\bar u}_k^j$ for all $u_k \in Z_j$. Note that $c_j$ denotes both a clause
of ${\cal C}$ and the corresponding vertex of $G$, and that both $u_k^j$ and ${\bar u}_k^j$ are symbols rather
than mathematical expressions.}
\vspace{-2mm}
\item[(7)] {\hspace{-1.5pt}}{For each pair $F_i, H_j$, a vertex $\sigma$ of $F_i$ is adjacent to every vertex of $H_j$ if the literal 
$\sigma$ occurs in the clause $c_j$; otherwise, vertex $\sigma$ is adjacent to no vertex of $H_j$, except to the vertex
$\sigma^j$ when $\sigma^*=u_i \in Z_j$.  Note that $\sigma^j=u_i^j$ and $\bar{\sigma}^j=\bar{u}_i^j$ 
if $\sigma=u_i$;  $\sigma^j=\bar{u}_i^j$ and $\bar{\sigma}^j=u_i^j$ if $\sigma=\bar{u}_i$.} 
\vspace{-2mm}
\item[(8)] {\hspace{-1.5pt}}{Finally, for each pair $H_i, H_j$, add all possible edges between $H_i$ and $H_j$. }  
\end{enumerate}
\vspace{-2mm}
\noindent This completes the construction of $G$; see Figure 1 for an illustration. It is easy to see that the construction can be accomplished 
in polynomial time, and the resulting graph contains $O(mn)$ vertices and $O(m^2n^2)$ edges. 

\begin{figure}[H]

\centering

\includegraphics[width=0.8\textwidth]{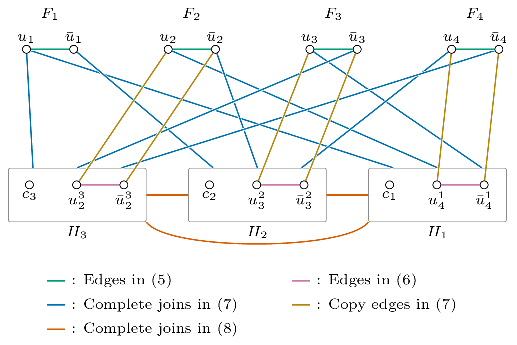}

\caption{The graph $G$ for the $\mathsf{3SAT}$ instance (in $\mathsf{CNF}$) with clauses $c_1=(u_1\vee u_2\vee u_3),\ c_2=(\bar u_1\vee\bar u_2\vee u_4),\ c_3=(u_1\vee\bar u_3\vee\bar u_4)$.
Each blue edge represents a complete join between a vertex in $F_i$ and all vertices in $H_j$ for some $i\in\{1,2,3,4\}$ and $j\in\{1,2,3\}$. For example, vertex $u_1$ is adjacent to every
vertex in $H_3$; vertex $\bar{u}_1$ is adjacent to every vertex in $H_2$; and vertex $u_3$ is adjacent to every vertex in $H_1$.}

\label{fig:construction}

\end{figure}

Let $V_1:=\cup_{i=1}^n \{u_i, {\bar u}_i\}$ and $V_2:=V_1 \cup \{c_1,c_2, \ldots, c_m\}$ be
two vertex subsets of $G$. For $i=1,2$, let $G_i$ be the subgraph of $G$ induced by $V_i$, and let
${\cal S}_i$ be the family of all maximal stable sets of $G_i$. Since each clause $c_j$ contains three distinct
literals, 
 
(9) vertex $c_j$ is adjacent to precisely three vertices of $G_1$ for $1\le j \le m$.  

For each $c_j$, let $N_1(c_j)$ denote the set of all neighbors of $c_j$ in $G_1$, and let $\bar{N}_1(c_j):=\{\bar{\sigma}: \sigma \in N_1(c_j)\}$.
For example, if $u_i \in N_1(c_j)$, then $\bar{u}_i \in \bar{N}_1(c_j)$; if $\bar{u}_i \in N_1(c_j)$, then $u_i \in \bar{N}_1(c_j)$.
By (9), we have $|N_1(c_j)|=3$.

From (3) and the construction of $G_1$ and $G_2$, it follows that

(10) A vertex subset $S\in {\cal S}_1$ if and only if it consists of precisely one vertex of $F_i$ for each $1\le i \le n$.  
A vertex subset $S\in {\cal S}_2 \de {\cal S}_1$ if and only if it is of the form $I\cup \{c_j\}$, where $1\le j \le m$
and $\bar{N}_1(c_j) \subseteq I\in {\cal S}_1$. Moreover, each vertex $c_j$ is contained in some $S\in 
{\cal S}_2 \de {\cal S}_1$ for $1\le j \le m$.

\vskip 2mm
We now explain the rationale behind the above construction. Given the structure of the \(\mathsf{3SAT}\) instance, it is natural to 
introduce the graphs $G_1$ and $G_2$ as above. By (1), each vertex $u$ of $G_1$ is contained in a maximal clique 
$C$ of $G_2$ such that $C\de \{u\} \subseteq \{c_1, c_2, \ldots, c_m\}$. This $C$ might be disjoint from 
some $S \in  {\cal S}_2 \de {\cal S}_1$. How should we tackle such a disjoint pair $C,S$?  We may appeal to the 
following technique: Let $J$ be obtained from $G_2$ by adding a new vertex $v$ and making it adjacent to all 
vertices in $C$ but to no vertex in $S$; the vertex $v$ may or may not be adjacent to vertices outside $C \cup S$. Then $S\cup \{v\}$ 
becomes a maximal stable set of $J$, which intersects every maximal clique of $J$ containing $C \cup \{v\}$. Thus, after adding $v$,
the disjoint pair $C, S$ is no longer problematic. 

There may, however, be many such disjoint pairs $C,S$ in $G_2$, with $c_j \in S \in {\cal S}_2 \de {\cal S}_1$. 
To eliminate all of them, we apply the same technique repeatedly. This process yields precisely the vertices
in $H_j \de \{c_j\}$ together with their corresponding edges.  
\vskip 2mm

Let ${\cal S}$ be the family of all maximal stable sets of $G$. The following statement says that its members fall into only two categories.

(11) A vertex subset $S\in {\cal S} \de {\cal S}_1$ if and only if it is of the form $I\cup \{c_j\} \cup \{\bar{\sigma}^j: \sigma \in I\de \bar{N}_1(c_j)\}$,
where $1\le j \le m$ and $\bar{N}_1(c_j) \subseteq I \in {\cal S}_1$. (Note that now $I\cup \{c_j\} \in {\cal S}_2 \de {\cal S}_1$ by (10),
$Z_j=\{\sigma^*: \sigma \in I\de \bar{N}_1(c_j)\}$, and $\{\bar{\sigma}^j: \sigma \in I\de \bar{N}_1(c_j)\}$ consists of precisely one vertex from each 
edge $u_k^j{\bar u}_k^j$ of $H_j\de \{c_j\}$, with $u_k \in Z_j$.)

To justify this, let $S\in {\cal S}\de {\cal S}_1$. Then $S$ contains a vertex $v$ from some $H_j$. By (8),  
$v$ is adjacent to all vertices of $H_i$ for any $i\ne j$. Hence $S$ contains no vertex from $H_i$ for any $i \ne j$,
as $S$ is a stable set.  By (7), every vertex in $N_1(c_j)$ is adjacent to every vertex in $H_j$. So $S \cap N_1(c_j) = \emptyset$.
Let $I=S \cap V_1$. Since $S$ is a maximal stable set, we must have $\bar{N}_1(c_j) \subseteq I\in {\cal S}_1$. In view of the structure
of $H_j$, we further obtain $c_j \in S$. Moreover, by definition, $Z_j=\{\sigma^*: \sigma \in I\de \bar{N}_1(c_j)\}$. By (7), each vertex
$\sigma$ in $I\de \bar{N}_1(c_j)$ is adjacent to exactly one vertex, namely $\sigma^j$, in $H_j$.  The maximality assumption on
$S$ then implies that $\{\bar{\sigma}^j: \sigma \in I\de \bar{N}_1(c_j)\} \subseteq S$. Combining the above observations, we 
conclude that $S=I\cup \{c_j\} \cup \{\bar{\sigma}^j: \sigma \in I\de \bar{N}_1(c_j)\}$.

Conversely, by (7) and (8), every set $S$ of the form $I\cup \{c_j\} \cup \{\bar{\sigma}^j: \sigma \in I\de \bar{N}_1(c_j)\}$,
with $1\le j \le m$ and $\bar{N}_1(c_j) \subseteq I \in {\cal S}_1$, is a stable set and dominates every vertex of $G$ outside
it. So $S$ is a maximal stable set of $G$, which belongs to ${\cal S} \de {\cal S}_1$ by (10). This proves (11).

(12) Each $S\in {\cal S} \de {\cal S}_1$ intersects each maximal clique of $G$.

Assume the contrary: $S \cap C=\emptyset$ for some $S\in {\cal S} \de {\cal S}_1$ and some maximal clique $C$ of $G$.  By (11),
we have $S= I\cup \{c_j\} \cup \{\bar{\sigma}^j: \sigma \in I\de \bar{N}_1(c_j)\}$,  with $1\le j \le m$ and $\bar{N}_1(c_j) \subseteq I \in {\cal S}_1$.
Since $C$ is disjoint from $S$, from (3) and the structure of $G_1$, we deduce that $C$ contains at most one vertex from $G_1$; let $\pi$ be this vertex, if it exists. Recall that 
$\pi \in \{u_i, {\bar u}_i\}$ for some $i$. Since $\bar{N}_1(c_j) \cup \{c_j\} \subseteq S$, we obtain $(\bar{N}_1(c_j) \cup \{c_j\}) \cap C=\emptyset$.
Hence

(13) either $\pi \in N_1(c_j)$ or $\pi \in \{\bar{\sigma}: \sigma \in I\de \bar{N}_1(c_j)\}$ (so variable $\pi^* \in Z_j$).  

Depending on whether $C$ is disjoint from $H_j$, we distinguish between two cases.

Case 1. $C$ contains no vertex of $H_j$.  

In this case, let $v:=\pi^j$ if $\pi$ exists and $\pi^* \in Z_j$, and let $v$ be an arbitrary vertex of $H_j\de \{c_j\}$
otherwise. By (7) and (13), $v$ is adjacent to $\pi$, if it exsits. Combining this with (8), we see that $v$ is adjacent 
to every vertex of $C$, contradicting the maximality assumption on $C$. 

Case 2. $C$ contains some vertex of $H_j$.  
   
In this case, $C$ contains a vertex $\omega^j$ from $H_j \de \{c_j\}$, where $\omega^j \in \{u_k^j, {\bar u}_k^j\}$ 
for some $u_k \in Z_j$, because $c_j \notin C$. By (11), we have $\bar{\omega}^j \in S$. If $\pi$ exists, then
either $\pi \in N_1(c_j)$ or $\pi \in \{\bar{\sigma}: \sigma \in I\de \bar{N}_1(c_j)\}$ (so variable $\pi^* \in Z_j$)  
by (13). Assume that $\pi \in N_1(c_j)$. Then $\bar{\omega}^j$ and $\pi$ are adjacent by (7). Combining this with (8), we conclude 
that $\bar{\omega}^j$ is adjacent to every vertex of $C$, contradicting the maximality assumption on $C$. If $\pi$ 
does not exist, we can reach a contradiction similarly. It remains to consider the subcase when $\pi$ exists and variable $\pi^* \in Z_j$. 
Since $\pi$ and $\omega^j$ are adjacent, $\pi=\omega$ by (7) and hence $\bar{\pi}= \bar{\omega}$. It follows from
(7) that $\bar{\omega}^j$ and $\bar{\pi}$ are also adjacent. Since $\pi \in C$, we have $\bar{\pi}\in I \subseteq S$. 
Hence both $\bar{\omega}^j$ and $\bar{\pi}$ are contained in the stable set $S$, contradicting its stability. So (12) holds.  

In view of (2) and the structure of $H_j$ for $1\le j \le m$, every vertex outside $\{c_1, c_2, \ldots, c_m\}$ has at least one neighbor
in this set. Therefore 

(14) the vertices $c_1, c_2, \ldots, c_m$ form a maximal clique $C$ of $G$.

Based on the above observations, we are ready to prove that $G$ is not CIS if and only if ${\cal C}$ is satisfiable. 

{\bf Sufficiency.} Suppose that $\tau: U\rightarrow \{\mbox{\em true, false}\}$ is a satisfying truth assignment for ${\cal C}$. 
Let $\sigma_i$ denote the literal in $\{u_i, {\bar u}_i\}$ such that $\tau(\sigma_i)=\mbox{\em true}$ for $1\le i \le n$. Let 
$S=\{\sigma_1, \sigma_2, \ldots, \sigma_n\}$ be the corresponding vertex subset of $G_1$. Note that $S\in {\cal S}_1$. Since each clause 
$c_j$ contains at least one true literal, vertex $c_j$ is adjacent to some vertex in $S$. Observe that  
$S$ is a maximal stable set of $G$, for otherwise, let $S'$ be a maximal stable set of $G$ containing $S$. By stability,
$c_j \notin S'$ for any $1\le j \le m$. So $S'$ contains at least one vertex of some $H_j \de \{c_j\}$ and hence $S'\in {\cal S} \de 
{\cal S}_1$, contradicting (11). Therefore the maximal stable set $S$ is disjoint from the maximal clique $C$ of $G$ exhibited in (14). 

{\bf Necessity.} Suppose that some maximal clique is disjoint from some maximal stable set $S$ of $G$. By (12),
we have $S \in {\cal S}_1$. Thus $S$ does not intersect the maximal clique $C$ exhibited in (14).
Let $\sigma_i$ denote the vertex in $\{u_i, {\bar u}_i\} \cap S$ for $1\le i \le n$. Define a truth assignment 
$\tau$ for ${\cal C}$, such that $\tau(\sigma_i)=\mbox{\em true}$ for each literal $\sigma_i$. Since $S$ is 
a maximal stable set, each vertex $c_j$ is adjacent to some vertex in $S$. Hence each clause $c_j$ is satisfied by $\tau$. 
Therefore $\tau: U\rightarrow \{\mbox{\em true, false}\}$ is a satisfying truth assignment for ${\cal C}$. This completes the proof of our theorem.

\end{document}